\documentclass[%
 reprint,
superscriptaddress,
 amsmath,amssymb,
prb,
]{revtex4-2}

\usepackage{graphicx}
\usepackage{dcolumn}
\usepackage{bm}
\usepackage{upgreek}

\begin{document}

\preprint{APS/123-QED}

\title{Spin-wave dispersion measurement by variable-gap propagating\\spin-wave spectroscopy}

\author{Marek Va\v{n}atka}
\email{marek.vanatka@live.com}
    \affiliation{CEITEC BUT, Brno University of Technology, 612 00 Brno, Czech Republic}%

\author{Krzysztof Szulc}
    \affiliation{ISQI, Faculty of Physics, Adam Mickiewicz University, Pozna\'{n}, Poland}
    
\author{Ond\v{r}ej Wojewoda}
    \affiliation{CEITEC BUT, Brno University of Technology, 612 00 Brno, Czech Republic}
    
\author{Carsten Dubs}
     \affiliation{INNOVENT e.V. Technologieentwicklung, 07745 Jena, Germany}

\author{Andrii Chumak}
     \affiliation{Faculty of Physics, University of Vienna, A-1090 Wien, Austria}

\author{Maciej Krawczyk}
    \affiliation{ISQI, Faculty of Physics, Adam Mickiewicz University, Pozna\'{n}, Poland}
    
\author{Oleksandr V. Dobrovolskiy}
     \affiliation{Faculty of Physics, University of Vienna, A-1090 Wien, Austria}

\author{Jaros\l{}aw W. K\l{}os}
    \affiliation{ISQI, Faculty of Physics, Adam Mickiewicz University, Pozna\'{n}, Poland}
    
\author{Michal  Urb\'{a}nek}
  \email{michal.urbanek@ceitec.vutbr.cz}
 \affiliation{CEITEC BUT, Brno University of Technology, 612 00 Brno, Czech Republic}%
 \affiliation{Institute of Physical Engineering, Brno University of Technology, 616 69 Brno, Czech Republic}

\date{\today}

\begin{abstract}
Magnonics is seen nowadays as a candidate technology for energy-efficient data processing in classical and quantum systems. Pronounced nonlinearity, anisotropy of dispersion relations and phase degree of freedom of spin waves require advanced methodology for probing spin waves at room as well as at mK temperatures. Yet, the use of the established optical techniques like Brillouin light scattering (BLS) or magneto optical Kerr effect (MOKE) at ultra-low temperatures is forbiddingly complicated. By contrast, microwave spectroscopy can be used at all temperatures but is usually lacking spatial and wavenumber resolution. Here, we develop a variable-gap propagating spin-wave spectroscopy (VG-PSWS) method for the deduction of the dispersion relation of spin waves in wide frequency and wavenumber range. The method is based on the phase-resolved analysis of the spin-wave transmission between two antennas with variable spacing, in conjunction with theoretical data treatment. We validate the method for the in-plane magnetized CoFeB and YIG thin films in $\bm{k}\perp \bm{B}$ and $\bm{k}\parallel \bm{B}$ geometries by deducing the full set of material and spin-wave parameters, including spin-wave dispersion, hybridization of the fundamental mode with the higher-order perpendicular standing spin-wave modes and surface spin pinning. The compatibility of microwaves with low temperatures makes this approach attractive for cryogenic magnonics at the nanoscale.

\end{abstract}

\maketitle


\section{\label{sec1}Introduction}

Properties of magnetic materials are of high interest due to several application concepts regarding, e.g., memories, sensors, microwave devices, or logic devices \cite{Dieny2020rev}. In the emerging field of magnonics, which utilizes spin-waves for data transport and processing, the essential system characteristic is the spin-wave dispersion relation. It provides a connection between the $k$-space and frequency space, and it also dictates other properties like the group velocity $v_g$ and decay length $\Lambda$. In thin films (approx. below 30\,nm), only the fundamental mode is observed when the experimentally accessible frequency range is limited to few GHz. In contrast, the spin-wave dispersion of thicker films can be rather complex as multiple perpendicularly standing modes may appear in the spectrum, exhibiting frequency crossing and hybridization \cite{tacchi_strongly_2019}. The spin-wave dispersion measurement is typically done by $k$-resolved \cite{tacchi_strongly_2019,mathieu_lateral_1998} or phase-resolved \cite{vogt_all-optical_2009,demidov_control_2009} Brillouin Light Scattering (BLS). Current interest in quantum computing, quantum magnonics \cite{lachance-quirion_entanglement-based_2020} and in superconductor/ferromagnet hybrid systems\cite{dobrovolskiy_magnonfluxon_2019,golovchanskiy_ferromagnet/superconductor_2018}, rises the need for material characterization at ultra-low temperatures, where optical access is typically extremely complicated. All electrical measurements are usually preferable in these applications.

The spin-wave dispersion measurement is also possible using the propagating spin-wave spectroscopy (PSWS). The PSWS is a technique which uses a vector network analyzer (VNA) connected to a pair of microwave antennas (e.g., striplines or coplanar waveguides) by microwave probes \cite{Baill2001,devolder_measuring_2021}. The two antennas (i.e., spin-wave transmitter and receiver) have a gap between them over which the spin-waves propagate, as shown in Fig. 1(a). Transmitting antenna is powered by the VNA’s microwave source and the receiving antenna serves as an induction pick-up detected by the VNA’s second port. The antenna type determines the excitation properties and can be adjusted to the experiment. Main three types of antennas used in experiments are striplines (rectangular wires), U-shaped ground-signal (GS) antennas, and coplanar waveguide (CPW) antennas. Schematic geometry of all three antenna types is shown in Fig. 1 (b). Striplines provide a continuous spectrum where the maximum excited $k$-vector is limited by the stripline width. Ciubotaru et al. \cite{ciubotaru_all_2016} showed scalability of the antennas, where 125\,nm wide striplines provided a wide continuous $k$-vector band. Good alternatives are GS   antennas, e.g. \cite{Baill2001,yamanoi_spin_2013,bhaskar_backward_2020}, or coplanar waveguides (CPW), e.g. \cite{bailleul_propagating_2003,vlaminck_spin-wave_2010,gruszecki_microwave_2016,qin_propagating_2018}, both providing a filtering capability for the $k$-vector spectrum allowing only specific ranges to exist. PSWS can be used on both nanostructured materials (stripes) \cite{bailleul_propagating_2003,vlaminck_spin-wave_2010,yamanoi_spin_2013,yu_magnetic_2015,ciubotaru_all_2016,collet_spin-wave_2017,bhaskar_backward_2020} as well as layers \cite{krysztofik_characterization_2017,qin_propagating_2018}. It was previously shown that the PSWS signal can be negligibly different for continuous layers and wide stripes \cite{qin_propagating_2018}. PSWS signals can also be modeled \cite{vlaminck_spin-wave_2010,qin_propagating_2018,sushruth_electrical_2020}. 

In previous reports, the spin-wave dispersion was extracted from PSWS spectra measured on yttrium iron garnet (YIG) using the CPW excitation. As the CPWs excitation spectrum exhibits distinct peaks in $k$-space, it allows extracting one point in spin-wave dispersion for each peak. The central $k$-vector of each peak is then assigned to a frequency from either the envelope of the $S_{21}$ sweep \cite{yu_magnetic_2015,chen_spin_2018} or by fitting the $S_{21}$ spectrum \cite{qin_propagating_2018}. This approach is limited to only several extracted points, and it is not easily transferable to metallic materials because of the low signal amplitude (compared to YIG) caused by large damping, making it impossible to use more than two peaks from the CPW antenna’s excitation spectrum. 

Here, we show that the spin-wave dispersion measurement using VNA is possible with a high level of details determined by the VNA frequency step.  

\section{\label{sec2}measurement setup and sample preparation}

Our setup uses Rohde \& Schwarz ZVA50 VNA and GGB industries microwave probes to establish a connection to the two antennas lithographically fabricated on top of CoFeB and YIG thin films. The lithography process consisted of e-beam patterning using PMMA resist, e-beam evaporation of Ti 5\,nm/Cu 85\,nm/Au 10\,nm multilayer, and lift-off. The CoFeB films (nominal thicknesses 30\,nm and 100\,nm) were magnetron-sputtered from Co$_{40}$Fe$_{40}$B$_{20}$ (at.\,\%) target on (100) GaAs substrate with 5\,nm Ta buffer layer. The (111) YIG films were grown by liquid phase epitaxy on top of a 500\,$\upmu$m, thick (111) gadolinium gallium garnet substrate \cite{dubs_sub-micrometer_2017,dubs_low_2020}. The samples were placed in a gap of a rotatable electromagnet allowing to apply an in-plane magnetic field up to 400\,mT in an arbitrary direction with respect to the spin-wave propagation direction. The VNA was set using a calibration substrate supplied with the microwave probes. A power sweep was performed before measuring each type of the sample-antenna combination to find a suitable power level that avoids nonlinear phenomena \cite{zakeri_spin_2007} and maintains a sufficient signal-to-noise ratio.

\begin{figure}[b]
\includegraphics{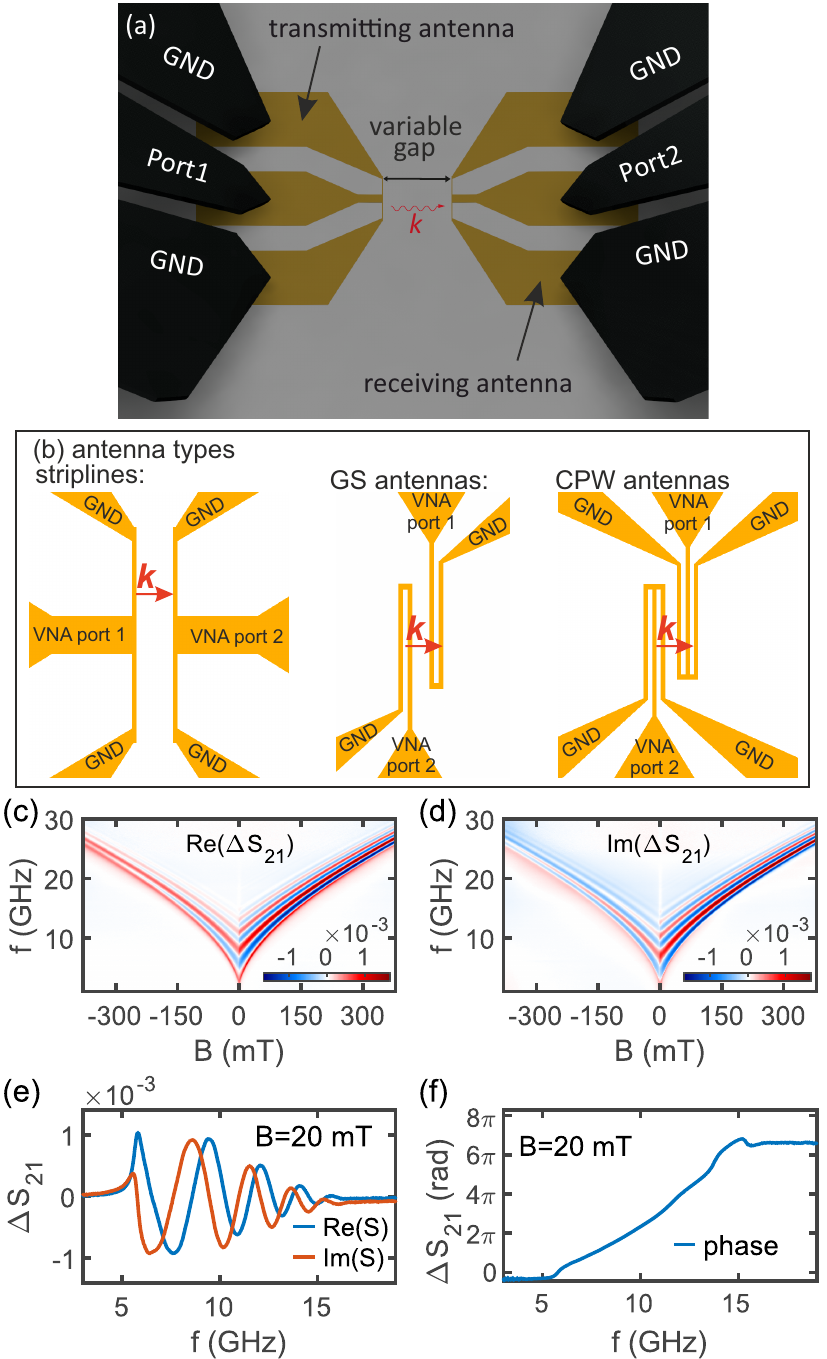}
\caption{\label{Fig1} (a) Schematics of the PSWS experiment using a pair of stripline antennas. (b) Schematic geometry of the three most commonly used antenna types. The outcomes of the PSWS measurement: (c) real and (d) imaginary part of $\Delta S_{21}(B,f)$  for 30\,nm thick CoFeB layer 30\,nm in $\bm{k}\perp \bm{B}$ geometry and the gap width of $1.8$\,$\upmu$m. The measurement shows the nonreciprocity of spin-wave propagation for the $\bm{k}\perp \bm{B}$ geometry which is reflected by the larger signal in $+B$ fields than in $-B$ fields. The plots of (e) show real and imaginary parts of $\Delta S_{21}(f)$, and (f) the unwrapped phase of $\Delta S_{21}(f)$, at fixed field $B=20$\,mT.}
\end{figure}

The VNA controls and analyses the electric signals of transmitting and receiving antennas both in the domain of amplitude and phase, therefore, it can measure the phase acquired by the spectral components of spin wave while it propagates between antennas. Our analysis is based on the $S_{21}$ transmission parameter. The transmitted spin-wave signal is modified by a nonmagnetic background $S_{21}^{\mathrm{background}}$ that is always present in the experiment due to direct electromagnetic crosstalk between the antennas. This background is constant for different values of static magnetic field, and therefore it is possible to evaluate it as the median over all measured magnetic fields. The subtracted signal $\Delta S_{21}$ is then calculated as: $\Delta S_{21}=S_{21}-S_{21}^{\mathrm{background}}$.

Fig. 1 (c,d) shows the $\Delta S_{21}$ signal, measured at 0\,dBm power output for a 30\,nm CoFeB thin film, over the gap 1.8\,$\upmu$m in the $\bm{k}\perp \bm{B}$ geometry (magnetostatic surface waves), and with 500\,nm wide striplines used as excitation and detection antennas. This geometry is known to be nonreciprocal with an exponential distribution of the dynamic magnetization along the layer’s thickness due to the surface localization of the mode \cite{schneider_phase_2008,sekiguchi_nonreciprocal_2010}. The higher signal amplitude in the $+B$ part of the spectrum is caused by both the stronger excitation and by the larger induction pick up from spin-waves propagating at the nearer surface. To achieve the best result, we can focus on $+B$ part of the spectrum (or alternatively on the $-B$ part and use the reverse transmission parameter $S_{21}$). Fig. 1(d) shows the plot of real and imaginary parts of $\Delta S_{21}$ measured at 20\,mT. The corresponding phase, which was unwrapped, is shown in Fig. 1(e). The phase rises from the ferromagnetic resonance (FMR) frequency up until it reaches the antenna’s excitation limits. Beyond this point, the signal loses its coherency due to insufficient signal-to-noise ratio, and therefore the phase stops evolving. The slope of the phase depends on the gap size – it changes more rapidly for wider gaps.

In the next step, we repeat the measurement on multiple instances of identical antenna structures (with varying gap width) prepared on the same 30\,nm CoFeB thin film sample. The phases measured over 11 gap widths are shown in Fig. 2(a). The phases are on the same level before the frequency reaches FMR, and then they start to rise. The lowest phase corresponds to the smallest gap width and the uppermost to the largest measured gap width. Then we project the measured phases into a phase – gap width plot [selected frequencies are plotted in Fig. 2(b)]. In this projection, the phase shows a linear dependence (for a coherent plane wave) that can be fitted; the slope equals to the $k$-vector at the given frequency. Now we can plot the extracted $k$-vectors against their frequencies, showing the resulting dispersion relation in Fig. 2(c). To confirm the result, we remeasured the same sample using phase-resolved BLS, and we found a very good agreement. The comparison of the dispersion relations measured by both techniques (VNA – red, BLS – blue) is shown in Fig. 2(c), and the comparison of the measured phase evolution at the frequency of 11.5\,GHz is shown in Fig. 2(d).

\begin{figure}[t]
\includegraphics{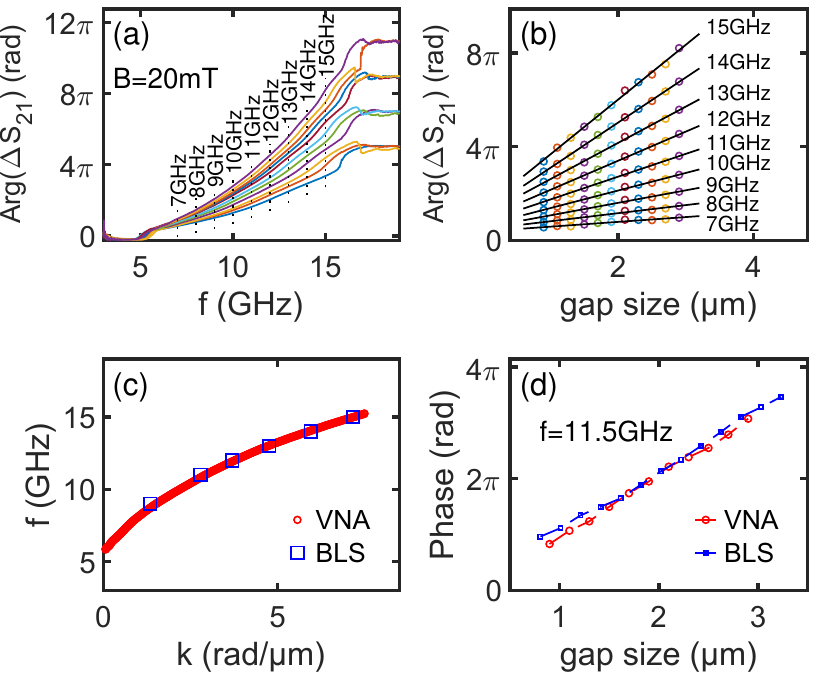}
\caption{\label{Fig2} Extraction of the dispersion relation from the outcomes of PSWS experiment on 30\,nm CoFeB thin film at $B=20$\,mT in $\bm{k}\perp \bm{B}$ geometry. (a) The unwrapped $\Delta S_{21}$ phases measured over gap widths ranging from $0.9$\,$\upmu$m (the lowest line) to $2.9$\,$\upmu$m (the steepest line) with the step of 200\,nm. (b) The representative fits of the phase at selected frequencies versus gap size where the slope of the fit yields the desired $k$-vector at that frequency. (c) The dispersion relation extracted for all frequencies within the range with sufficient PSWS signal (red line). The dispersion obtained by phase-resolved BLS is plotted in the blue points for comparison. (d) The compared outcomes of the phase measurements for VNA (red circles) and phase-resolved BLS (blue squares) at single frequency $f=11.5$\,GHz.}
\end{figure}

\section{\label{sec3}measured spin-wave dispersion relations}

Fig. 3(a) shows dispersion relations of the same 30\,nm CoFeB thin film measured in $\bm{k}\perp \bm{B}$  geometry in magnetic fields ranging from 20\,mT to 380\,mT with a step of 60\,mT. The sample was also measured in $\bm{k}\parallel \bm{B}$ geometry, but the measured signal was insufficient to reconstruct the dispersion relation. For all measured fields, it was possible to evaluate the dispersion for $k$-vectors ranging from 0 up to 8\,rad/µm. The upper limit is given by the excitation efficiency \cite{vlaminck_spin-wave_2010} of the used antenna (see blue lines in Fig. 3). By fitting the measured dispersions using the Kalinikos-Slavin model \cite{kalinikos_theory_1986}, we were able to obtain material parameters of the measured thin films (see black lines in Fig. 3 for the model fits and the figure caption for the fitting results).

In addition to the 30\,nm CoFeB thin film, we also measured 100\,nm thick CoFeB [Fig. 3(b)] and 100\,nm thick YIG [Fig. 3(c)] films to further explore the possibilities of the presented technique. The measurements were performed using different antenna types (stripline, GS, CPW) to see their influence on the quality of the obtained dispersions, and they were also evaluated for multiple magnetic fields in $\bm{k}\perp \bm{B}$ and in $\bm{k}\parallel \bm{B}$ geometries. The 100\,nm thick CoFeB film was measured in the fields ranging from 20 to 380\,mT with a step of 60\,mT at a power of 0\,dBm. In Fig. 3(b) we show the dispersion measured using 500\,nm wide stripline antenna for $\bm{k}\perp \bm{B}$ geometry and with 500\,nm coplanar waveguide (signal and ground line widths, as well as signal-to-ground gap, were 500\,nm) in $\bm{k}\perp \bm{B}$ geometry. We were able to obtain a dispersion in $\bm{k}\parallel \bm{B}$ geometry also using a 500\,nm wide stripline antenna but with substantially worse quality due to its lower excitation efficiency. The 100\,nm thick YIG film was measured in $\bm{k}\perp \bm{B}$ and $\bm{k}\parallel \bm{B}$ geometries in the fields ranging from 20 to 200\,mT with a step of 20\,mT at a power of  $-30$\,dBm. In this case, it was possible to obtain a dispersion in both geometries using all types of antennas. Fig. 3(c) shows data acquired by GS  antennas with gap widths from 1.0 to 3.4\,$\upmu$m with 400\,nm step. As in the case of the coplanar waveguide, the dimensions of the GS  antennas were 500\,nm (signal and ground line widths and signal-to-ground gap width). The dispersions are plotted for magnetic fields from 20 to 200\,mT with a step of 20\,mT and the power output was set to $-30$\,dBm.

\begin{figure*}[t]
\includegraphics{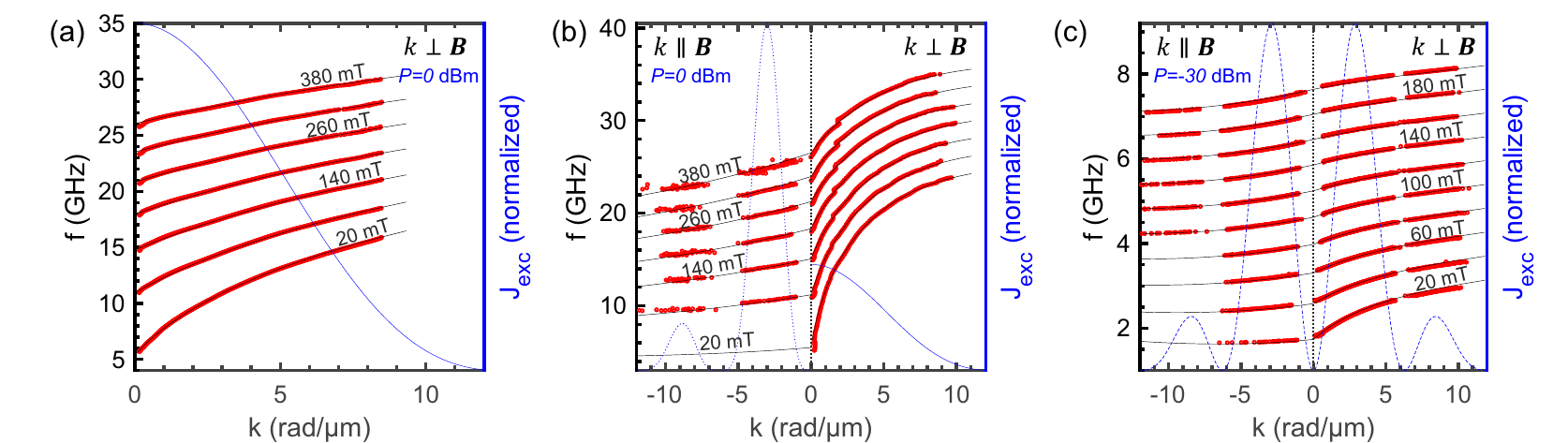}
\caption{\label{Fig3} The dispersion relations measured using the PSWS technique (red points) compared to the analytic dispersions from Kalinikos-Slavin model (black lines) for (a) 30\,nm thick CoFeB (b) 100\,nm thick CoFeB, (c) 100\,nm thick YIG layers. For clarity, the data for $\bm{k}\perp \bm{B}$ are plotted in the domain of positive wave numbers $k$ and the $\bm{k}\parallel \bm{B}$ data are presented with negative $k$. The samples were excited by stripline, CPW, and GS antennas with all lateral wire widths dimensions of 500\,nm. The blue lines show calculated (and normalized) wave number excitation spectra of the antennas, calculated form the spatial distributions of microwave field (the solid, dashed and dotted lines correspond to the spectra of stripline, GS, and CPW antenna, respectively). For Kalinikos-Slavin model we fitted the following values of parameters: (a) $M_s=1230$\,kA/m ($\mu_0M_s=1.54$\,T), $\gamma/2\pi=30.5$\,GHz/T, $t=29.0$\,nm; (b) $M_s=1310$\,kA/m ($\mu_0M_s=1.64$\, T), $\gamma/2\pi=30.1$\,GHz/T; (c) $M_s=$\,133 kA/m ($\mu_0M_s=0.166$\,T), $\gamma/2\pi=28.3$\,GHz/T, where $t$ stands for thickness of the layer. In fitting procedure, we  fixed the exchange stiffness to $A_\mathrm{ex}=15$\,pJ/m$^3$ for CoFeB films and $A_\mathrm{ex}=3.6$\,pJ/m$^3$ for the YIG film.}
\end{figure*}

In the dispersion measured on 100\,nm thick CoFeB film, we further identified visible hybridizations at the crossings’ positions between the $n=0$ fundamental spin-wave mode and $n=1,2,3$ higher-order perpendicularly standing spin-wave modes (see Fig. 4). The hybridizations are present in the measured dispersion relation as gap openings, with a portion of the measured data evolving towards smaller $k$-vectors [see Fig. 4(d) for the most prominent example]. This part of the dispersion has no physical meaning and is only a result of the data processing described earlier in the text. The hybridizations are also directly visible as dips in the magnitude $\mathrm{Abs}(\Delta S_{21})$ of the transmission spectra [see Fig. 4(b)]. 

\section{\label{sec4}numerical modeling}

To investigate numerically the hybridization between the fundamental mode and perpendicularly standing modes, we solve Landau-Lifshitz equation (LLE) using the finite-element method (FEM) in the frequency domain. The LLE was linearized and implemented in FEM solver (COMSOL Multiphysics) as a set of differential equations for in-plane and the out-of-plane components of magnetization $m_\parallel$ and $m_\perp$:
\begin{equation}
    i \omega m_\parallel= |\gamma|\mu_0 [ (H_0-\frac{2A_\mathrm{ex}}{\mu_0 M_s}\Delta  ) m_\perp + M_s \partial_{x_\perp} \varphi],
\end{equation}
\begin{equation}
    i \omega m_\perp= |\gamma|\mu_0 [ (H_0-\frac{2A_\mathrm{ex}}{\mu_0 M_s}\Delta  ) m_\parallel + M_s \partial_{x_\parallel} \varphi],
\end{equation}
together with the equation for the scalar magnetostatic potential $\varphi$ derived from Maxwell equations in magnetostatic approximation \cite{noauthor_spin_2009,rychly_spin_2017}:
\begin{equation}
   \Delta \varphi-\partial_{x_\parallel}m_\parallel-\partial_{x_\perp}m_\perp=0,
\end{equation}
where $x_\parallel$ and $x_\perp$ denote the directions of the corresponding dynamic components of magnetization.

In contrast to the analytical model of Kalinikos and Slavin, our numerical calculations truly reproduce the spin dispersion relation in the crossover dipolar-exchange regime \cite{tacchi_strongly_2019}, including the hybridizations between the fundamental spin-wave mode and perpendicularly standing spin-wave modes. In our model, we also consider the presence of surface anisotropy $K_s$. The anisotropy is expected to be different on the bottom and the top faces of the magnetic layer interfaced with different materials. The surface anisotropy and its asymmetry (between the top and bottom face) is responsible for the spin-wave pinning and the strength of the hybridization between fundamental and perpendicularly standing modes. The pinning is implemented in the boundary conditions \cite{rado_spin-wave_1959}
\begin{equation}
  \partial_{x_\perp}m_\parallel=0,
\end{equation}
\begin{equation}
   A_\mathrm{ex} \partial_{x_\perp}m_\perp - K_s m_\perp=0.
\end{equation}

The numerical model (as well as the analytical model by Kalinikos and Slavin) is characterized by a few material parameters. We fixed the value of exchange stiffness to $A_\mathrm{ex}=14$\,pJ/m$^2$  and layer’s thickness to nominal value $d=100$\,nm. The values of saturation magnetization $M_s=1275$\,kA/m and gyromagnetic ratio $|\gamma|/2\pi=30.8$\,GHz/T were selected to fit the ferromagnetic resonance frequency (i.e., the frequency of fundamental mode at $k = 0$) and the slope of the dispersion for fundamental mode. The surface anisotropy $K_s$ determines the spin-wave pinning and therefore is important both for the quantization (and the frequencies) of the perpendicularly standing modes and the strength of the hybridization between  and perpendicularly standing modes. To obtain the proper values of the frequencies of perpendicularly standing modes, we did not need to reduce the thickness below the nominal value 100\,nm but we had to introduce the non-zero $K_s$ instead, which seems to be a more realistic approach. The experimental data show that the hybridizations of fundamental mode with the perpendicularly standing mode are observed both for the perpendicularly standing modes quantized with even ($n=2$) and odd number of nodes ($n = 1, 3$) across the layer. The effect for $n = 1, 3$ is hardly visible in dispersion relation [Fig. 4(c,e)] but is quite distinctive in the measurement of the transmission amplitude [Fig. 4(b)]. These results indicate that the asymmetric pinning (and different values of surface anisotropy on both faces of the CoFeB layer) must be considered to obtain less symmetric profiles of spin-wave modes across the thickness of the layer, which in turn, gives the non-zero cross-sections between fundamental mode and perpendicularly standing modes of odd numbers of nodes. There is always some degree of freedom for choosing the values of $K_s$ on both faces, therefore, we decided to consider the simplest case where $K_s=0$ and 1600\,$\upmu$J/m$^2$ at the top (interface with vacuum) and bottom (interface with GaAs). With these values, we succeeded in the induction of the hybridization of fundamental mode ($n=0$) with the first ($n = 1$) and third ($n = 3$) perpendicularly standing modes, characterized by the width comparable to the widths of the deeps of the transmission $\Delta S_{12}$ [orange stripes in Fig. 4(a,b)]. It is worth noting that in the case of symmetric pinning (we took $K_s=700$\,$\upmu$J/m$^2$  on both faces), we do not observe the hybridization with the first ($n = 1$) and third ($n = 3$) perpendicularly standing modes [see blue dashed lines in Fig. 4(c,e)].

\begin{figure}
\includegraphics{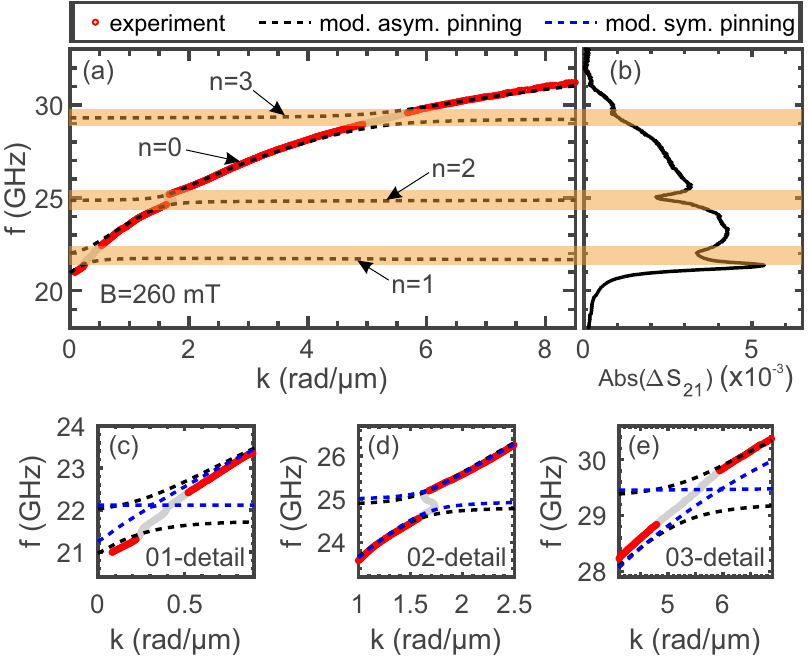}
\caption{\label{Fig4} (a) Measured dispersion relation (red points) for 100\,nm thick CoFeB layer at 260\,nm exhibiting hybridized spin-wave modes, visible mainly at the $n = 02$ crossing. The dashed line represents dispersion simulation with asymmetrical surface pinning conditions with parameters $K_S=0$ and $1600$\,$\upmu$J/m$^2$  on top and bottom face of the layer, respectively. (b) Corresponding magnitude of the $\Delta S_{21}$ transmission parameter exhibiting dips at the crossings’ positions ([$\mathrm{Abs}(\Delta S_{21})$] data shown for the gap width of 1.8\,$\upmu$m). (c-e) details of the 01, 02, and 03 crossings, respectively. Dashed blue lines correspond to the case of symmetric pinning ($K_S=700$\,$\upmu$J/m$^2$ on both faces).}
\end{figure}

\section{\label{sec5}additional data analysis and discussion}

From the measured dispersion relations, it is possible to extract also other parameters important for spin-wave propagation. The group velocity $v_g$ can be calculated as the numerical derivative of the dispersion $v_g=2\pi \mathrm{d}f/\mathrm{d}k$ and the lifetime $\tau$ can be obtained as the numerical derivative of field-dependent dispersion \cite{noauthor_spin_2009}: $\tau=(\alpha \omega \frac{\partial\omega}{\partial\omega_B})^{-1}=(\frac{\alpha 4 \pi^2 f}{\gamma} \frac{\partial f}{\partial B})^{-1}$, where  $\omega_B=\gamma B$ and $\alpha$ is a damping constant.

The decay length $\delta$ can be then calculated by multiplying the group velocity and the lifetime $\delta=v_g \tau$. In Fig. 5(a), we plot the decay length obtained by the above-described procedure (orange line) and compare it with the decay length obtained using the more traditional approach, i.e., by fitting the exponential decay of the magnitude of $\Delta S_{21}$ signal (blue line). Here, the decay length $\delta$ was obtained by fitting [see Fig. 5(b) for representative fits] the formula $\mathrm{Abs}(\Delta S_{21})=Ie^{-g/\delta}$ in logarithmic form $\mathrm{ln}[\mathrm{Abs}(\Delta S_{21})]=\mathrm{ln}(I)-\frac{1}{\delta} g$ , where $g$ is the gap distance, and $I$ is a free parameter proportional to signal strength. In Fig. 5(c), we plot field-dependent lifetime evaluated using the decay length obtained from the derivative of field-dependent dispersion ($\partial\omega/\partial\omega_B$, orange circles) and from fitting the exponential decay of the magnitude of $\Delta S_{21}$ signal (blue diamonds). Both approaches give roughly the same results for both the decay length and the lifetime, but the decay lengths obtained using the $\partial\omega/\partial\omega_B$ approach agree better with the analytical model [see Fig. 5(a), green line]. The lower quality of the data obtained from fitting the exponential decay of the magnitude of $\Delta S_{21}$  signal is caused by the limitation of our experimental arrangement. Due to the finite length of the excitation antenna, spin-wave caustics can form \cite{schneider_nondiffractive_2010}. An example of such caustics measured by BLS microscopy is shown in Fig. 5(e). The phase in the caustic beam is spatially incoherent \cite{korner_excitation_2017} and the focusing effect \cite{veerakumar_magnon_2006} causes modulation of the spin-wave intensity along the propagation direction. The associated adverse effects can be avoided by measuring at propagation distances that are sufficiently small with respect to the stripline length. In our experiments, we used the stripline length of 10\,$\upmu$m, and the maximum measured propagation distance was 2.9\,$\upmu$m. For phase-resolved measurement such as evaluation of dispersions, the short propagation distance was sufficient. On the other hand, the change in signal at 2.9\,$\upmu$m propagation distance was not sufficient to obtain a fully reliable fit of the exponential decay of the $\Delta S_{21}$ magnitude and at longer propagation distances the measurements were distorted by caustics.

We also analyzed how the number of measured propagation distances (gaps) affected the reliability of the obtained dispersion. We took the data measured for 11 gap widths in total and then used different combinations of a reduced number of gap widths (starting from 2 up to 10) to evaluate the dispersion. The dispersion obtained from the reduced number of gap widths was then compared to the analytical fit Kalinikos-Slavin model \cite{kalinikos_theory_1986} of the dispersion obtained from fitting of the complete set of 11 measured points. As shown in Fig. 5(d), the mean frequency difference from the reference dispersion can rise to the maximum of 400\,MHz when using a combination of just two gap widths. This maximum quickly decreases to 200\,MHz for five gap widths and then stays around 100\,MHz for combinations of six and more gaps.

\begin{figure}
\includegraphics{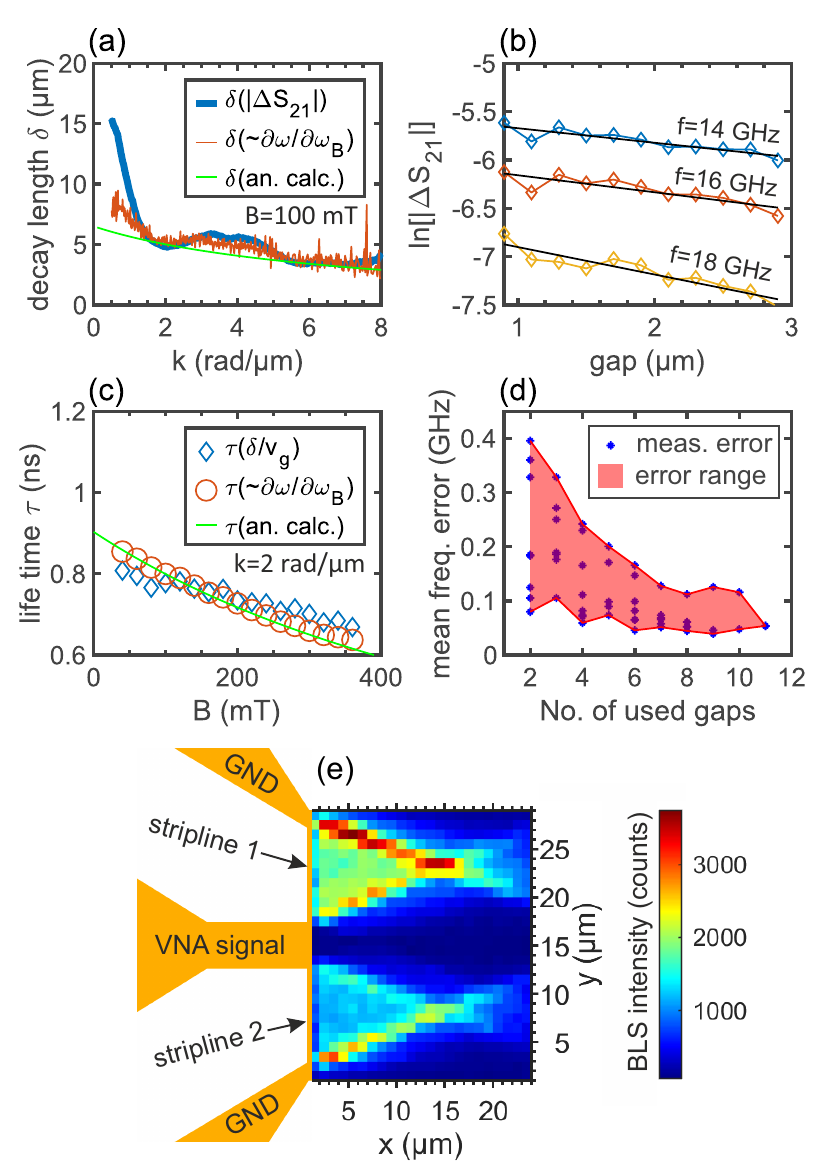}
\caption{\label{Fig5} The determination of decay length and lifetime from VG-PSWS measurement of 30\,nm thick CoFeB film (a-c,e) and the impact of the number of gaps on the quality of the result (d). (a) Blue line shows the spin-wave decay lengths obtained directly from exponential fits, orange line shows the decay length calculated by multiplication of the life time and the group velocity and the green line is calculated from the Kalinikos-Slavin model. (b) Representative exponential fits of the $\Delta S_{21}$ magnitude which were used to obtain the spin-wave propagation length. (c) Spin-wave life time extracted from the derivative of field-dependent dispersion using $\alpha=0.075$ (orange circles), calculated by dividing the propagation length (from $\Delta S_{21}$ magnitude) by the numerically calculated group velocity (blue diamonds) and calculated from the Kalinikos-Slavin model (green line). (d) Mean frequency difference of the dispersion obtained from the reduced number of gap widths compared to the analytical fit of the dispersion obtained from the complete set of 11 measured points. Blue points represent different combinations of gaps and the red area represents the difference spread. (e) Spin-wave caustics caused by the finite length of the antenna, measured on 100\,nm CoFeB film by BLS with exciation frequency of 11\,GHz in the external magnetic field of 26\,mT. Caustics can destroy the phase coherence of propagating spin-waves. Note that all presented data were collected at propagation distances (gap widths) between 0.9\,$\upmu$m to 2.9\,$\upmu$m to avoid the negative effects of the caustics formation.}
\end{figure}

As can be seen from the presented data, the variable-gap approach allows the reconstruction of the spin-wave dispersion relation, including detailed features that might be used for analysis of the spin-wave system. In order to obtain the highest possible data quality, we need to fabricate multiple pairs of antennas where the gap width and the step in the gap width (an increase of the gap between two pairs of antennas) must be optimized for each experiment. Note that this approach does not require precise knowledge of the excitation spot location (phase origin) because it does not affect the fit’s slope. It is only essential to know the relative differences between the propagation distances, i.e., the gap-width step, which can be easily measured (and precisely fabricated using e-beam lithography). 

The antenna design is sample-specific and needs to be tailored to fulfil experimental requirements. The antenna’s type and geometry must be able to excite the expected $k$-vector range with sufficient efficiency.   

The gap widths need to be in the optimum range considering the caustics formation and the spin-wave decay length for the given material and geometry. For example, for CoFeB 30\,nm in $\bm{k}\parallel \bm{B}$ geometry at $k=5$\,rad/$\upmu$m, we calculated decay lengths of 0.03\,$\upmu$m and 0.45\,$\upmu$m for magnetic fields of 20 and 200\,mT, respectively. Here, the quality of the measured data, even for the smallest fabricated gap distance of 1\,$\upmu$m, was not sufficient to evaluate dispersion; thus, this geometry is not presented in Fig. 3(a).

\begin{table*}[t]
\caption{\label{tab1}Comparison of experimental techniques used for measurement of spin-wave dispersion relations.}
\begin{ruledtabular}
\begin{tabular}{lcccccc}
  & Lateral  &	Resolution 	& Resolution &	Phase  & max $k$	& Experimental \\
   & resolution  &	 in $k$ (rad/$\upmu$m)	&  in $f$ (Hz) & extraction & (rad/$\upmu$m)	& geometries \\
    &  in (nm) &	& &  &	& (MS, BV, FV) \\
\hline
VG-PSWS [this work] &	-	& $<0.1$* &	1 &	yes &	10*** &	MS, BV, (FV possible)\\
\hline
PSWS \cite{qin_propagating_2018,sushruth_electrical_2020} &	- &	2*** &	1 &	no &	10*** &	MS, BV, FV \\
\hline
MOKE \cite{dreyer_spin-wave_2021,qin_nanoscale_2021} &	500 &	$< 0.1$* &	$1\cdot10^{8}$ ** &	yes &	10 &	BV, MS \\
\hline
Conventional BLS \cite{sebastian_micro-focused_2015,tacchi_strongly_2019} &	10,000 &	0.5 &	$1\cdot10^{8}$  &	no &	23.6 &	BV, MS \\ 
\hline
Phase-resolved BLS \cite{vogt_all-optical_2009, flajsman_zero-field_2020, wojewoda_propagation_2020} &	250 &	$< 0.1$* &	$1\cdot10^{8}$ ** &	yes &	7 &	BV, MS \\
\hline
STXM \cite{gros_nanoscale_2019,forster_nanoscale_2019} &	20 &	$< 0.1$* &	$1\cdot10^{8}$ ** &	yes &	30 &	BV, MS \\
\end{tabular}
\end{ruledtabular}
\begin{tabular}{l}
MS - magnetostatic surface spin waves, BV - backward volume spin waves, FV - forward volume spin waves\\
 * The $k$-resolution can be tailored to the magnetic system under study by an appropriate design of the experiment geometry. \\
 **The resolution in $f$ in these techniques is limited by the acquisition time.\\ 
 *** Excitation limited.
\end{tabular}

\end{table*}

Before fitting the data as plotted in Fig. 2(b), the phase needs to be correctly unwrapped. In the case of stripline antennas with a continuous excitation spectrum, it is possible to achieve correct unwrap even when the phases for neighboring gap widths at the same frequency are higher than $\pi$ rad by unwrapping the phase in the frequency spectrum [Fig. 2(a)]. It is because the frequency step size of the VNA is usually small and the phase shift of the neighboring points is always smaller than $\pi$ rad. On the other hand, the safest approach is to unwrap the phases when plotted against the gap size [Fig. 2(b)]. In this case, the phase change of neighboring points must be smaller than $\pi$ rad. This is necessary for antenna types with discrete excitation spectra (i.e., CPW, GS, ladders \cite{bang_excitation_2018} or meanders \cite{lucassen_optimizing_2019}). 

The step in the gap width defines the maximum $k$-vector for which the dispersion can be measured. The phase change of the spin wave along the step distance has to be smaller than $\pi$ rad except for certain cases discussed later. On the other hand, small $k$-vectors may have a phase change that is too small over a short step in the gap width. If we require an accurate fitting of small $k$-vectors, the step in the gap width should be large enough to provide sufficient phase change (we suggest at least 1\,rad) while respecting the decay length.

\section{\label{sec6}method comparison}

In Table 1, we compare the variable gap propagating spin-wave spectroscopy with other experimental techniques used to obtain spin-wave dispersion relations. To get detailed dispersion, one needs to have high resolution in both the $k$-vector and the frequency. Optical and X-ray techniques have low frequency resolution, which is typically limited to hundreds of MHz. In the case of conventional BLS, the resolution is determined directly by the Fabry-Perot interferometer \cite{scarponi_high-performance_2017}. The optical techniques using microwave excitation, i.e., Magneto-Optical Kerr Effect microscopy (MOKE), Phase-resolved BLS and Scanning Transmission X-ray Microscopy (STXM), are limited by slow signal acquisition.  It does not allow capturing the required span of frequencies with a high resolution in a reasonable time. The propagating spin-wave spectroscopy technique, which uses known positions of the excitation peaks of the CPW antenna, can capture the data with a high resolution in frequency. However, the $k$-resolution is limited with the finite widths of peaks in the excitation spectra of the used CPW antennas which in turn affect also the frequency resolution.    

 Big advantage of the VG-PSWS technique is that it does not require direct optical access to the sample, making the method very suitable for, e.g., experiments at ultra-low temperatures. In addition, compared to other techniques for spin-wave dispersion relation measurement VG-PSWS can achieve high resolution in frequency and $k$-vector at the same time. This combination allows capturing smooth dispersion curves over the span of all accessible $k$-vectors and fast acquisition times allow repeating the experiment in multiple magnetic field strengths and orientations. The full set of field-dependent dispersion curves measured for both $\bm{k}\perp \bm{B}$ and in $\bm{k}\parallel \bm{B}$ geometries present a robust 4-dimensional ($f$, $k$, $B$, angle) dataset that can be further evaluated, and all essential material and spin-wave parameters can be extracted from it.

A disadvantage of this method is the need for a set of antennas with multiple gap widths on top of the sample.  The other techniques can obtain the dispersion either without any antenna (conventional BLS), with one excitation antenna (MOKE, Phase-resolved BLS, STXM) or with a pair of excitation and detection antennas (PSWS). However, the need for multiple antennas may be overcome in the future by using freestanding positionable antennas \cite{hache_freestanding_2020}.

\section{\label{sec7}conclusion}

In conclusion, we presented a new method of extraction of high-quality spin-wave dispersion relation from propagating spin-wave spectroscopy (PSWS) measurements performed over several propagation distances. We demonstrated this technique on CoFeB and YIG thin films measured in $\bm{k}\perp \bm{B}$ and $\bm{k}\parallel \bm{B}$ geometries. The results on CoFeB thin film were verified by phase-resolved BLS measurement showing good agreement. When compared with the phase-resolved BLS, the VNA-based method provides more frequency measurement points in a shorter acquisition time. Fine detail measurement capability was demonstrated on the measurement and analysis of hybridized modes acquired on 100\,nm thick CoFeB thin film, revealing asymmetric surface pinning and the values of pinning parameters on both interfaces of the magnetic layer. The all electric nature of this method makes it very suitable for characterization of cryogenic and quantum magnonics systems and materials.

\begin{acknowledgements}
The work was supported by MEYS CR (project CZ.02.2.69/0.0/0.0/19\_073/0016948). CzechNanoLab project LM2018110 is gratefully acknowledged for the financial support of the measurements and sample fabrication at CEITEC Nano Research Infrastructure. 
O.W. was supported by Brno PhD talent scholarship.
J.W.K. and M.K. acknowledge the support of  the  National Science Centre - Poland for the projects UMO-2020/37/B/ST3/03936 and UMO-2020/39/O/ST5/02110.
O.V.D. acknowledges the Austrian Science Fund (FWF) for support through Grant No. I 4889 (CurviMag).
C.D. gratefully acknowledges financial support from the Deutsche Forschungsgemeinschaft (DFG, German Research Foundation) - 271741898.
\end{acknowledgements}

\bibliography{apssamp}

\end{document}